\documentclass[twocolumn,superscriptaddress,pr]{revtex4-1}
\usepackage{verbatim}
\usepackage{amsmath,amssymb}
\usepackage{graphicx}
\usepackage{color}
\usepackage[colorlinks,bookmarks=false,citecolor=blue,linkcolor=red,urlcolor=blue]{hyperref}
\usepackage{natbib}
\usepackage{times}

\DeclareMathOperator{\tr}{tr}

\begin{document}

\title{Entanglement negativity via replica trick: a Quantum Monte 
Carlo approach}

\author{Chia-Min Chung}
\affiliation{Department of Physics and Frontier Research Center on Fundamental and Applied 
Sciences of Matters, National Tsing Hua University, Hsinchu 30013, Taiwan}
\author{Vincenzo Alba}
\affiliation{Department of Physics and Arnold Sommerfeld
Center for Theoretical Physics, Ludwig-Maximilians-Universit\"at
M\"unchen, D-80333 M\"unchen, Germany}
\author{Lars Bonnes}
\affiliation{Institute for Theoretical Physics, University 
of Innsbruck, A-6020 Innsbruck, Austria}
\author{Pochung Chen}
\affiliation{Department of Physics and Frontier Research Center on Fundamental and Applied 
Sciences of Matters, National Tsing Hua University, Hsinchu 30013, Taiwan}
\affiliation{Physics Division, National Center for Theoretical Sciences, Hsinchu 30013, Taiwan}
\author{Andreas M. L\"{a}uchli}
\affiliation{Institute for Theoretical Physics, University of 
Innsbruck, A-6020 Innsbruck, Austria}

\date{\today}

\begin{abstract}
Motivated by recent developments in conformal field theory (CFT), we devise  
a Quantum Monte Carlo (QMC) method to calculate  the moments of the 
partially transposed reduced density matrix at finite temperature. 
These are used to construct  scale invariant combinations that are related to the 
{\it negativity}, a true measure of entanglement for two intervals embedded in a chain.
These quantities can serve as witnesses of criticality.
In particular, we study several scale invariant combinations of the moments 
for the 1D hard-core boson model. 
For two adjacent intervals unusual finite size corrections are present, 
showing parity effects that oscillate with a filling dependent period. These 
are more pronounced in the presence of boundaries. For large chains we find 
perfect agreement with CFT. Oppositely, for disjoint intervals corrections are 
more severe and CFT is recovered only asymptotically. Furthermore, we provide evidence 
that their exponent is the same as that  governing the corrections of the mutual 
information. Additionally we study the 1D Bose-Hubbard model in the superfluid phase.
Remarkably,  the finite-size effects are smaller and QMC data are already in impressive agreement 
with CFT at moderate large sizes.
\end{abstract}

%\pacs{}
\maketitle

\paragraph*{Introduction.---} 

The quest for universality  has long been a driving research theme at 
the border between condensed matter and quantum field theory. Recently,    
much progress has been achieved due to the deep relation between conformal 
field theory (CFT) and quantum entanglement~\cite{
osterloh2002,holzhey1994,vidal2003,calabrese2004}. 

Given a bipartition of a system (in a pure state $|\psi\rangle$) into 
two parts $A$ and $B$, a measure of their mutual entanglement is 
the von Neumann entropy $S_1\equiv\tr\rho_A\log\rho_A$, with $\rho_A
\equiv\tr_B|\psi\rangle\langle\psi|$ the reduced density matrix for 
$A$. Alternatively, the so-called Renyi entropy $S_{n}\equiv-1/(n-1)
\log\tr\rho_A^{n}$~\cite{amico2008,eisert2010,calabrese2009c} are 
also valid entanglement measures. 

It is now well established that the entropies contain {\it universal} 
information about 1D critical systems, namely the central charge~\cite{
holzhey1994,vidal2003,calabrese2004,cardy2010a} of the underlying CFT. 
Moreover, if subsystem $A$ consists of two (or many) disjoint intervals, as 
$A\equiv A_1\cup A_2$, the mutual information, $I_{A_1:A_2}\equiv S_{A_1}+
S_{A_2}-S_{A_1\cup A_2}$, depends on the full operator content of 
a  CFT~\cite{caraglio2008,furukawa2009,calabrese2009a,calabrese2009b,
calabrese2009c,igloi2010,fagotti2010,fagotti2011,fagotti2012,alba2010,
alba2011,calabrese2011,coser2013}. 

However, as the subsystem $A_1\cup A_2$ is generally in a mixed state, the mutual information is not a measure of their entanglement but of all (quantum and classical) correlations between $A_1$ and $A_2$~\cite{wolf08}.
Their entanglement, instead, 
can be quantified via the logarithmic {\it negativity} ${\cal E}$~\cite{
vidal2002}
\begin{equation}
{\cal E}\equiv\log||\rho_A^{T_2}||=\log\tr|\rho_A^{T_2}|.
\end{equation}
Here $\rho_A^{T_2}$ is the partially transposed reduced density matrix 
with respect to  $A_2$  (formally $\langle\varphi_1\varphi_2|\rho_A^{T_2}
|\varphi_1'\varphi_2'\rangle\equiv\langle\varphi_1\varphi'_2|\rho_A|
\varphi_1'\varphi_2\rangle$,  with $\{\varphi_1\}$, $\{\varphi_2\}$ 
being  a basis for $A_1,A_2$). 

Unlike the entropy, which contains non universal contributions, the negativity ${\cal E}$ 
is {\it fully} universal at a quantum critical point and, therefore, 
a useful tool to distinguish between different universality classes. 
This was originally argued on the basis of DMRG calculations~\cite{wichterich2009,
wichterich2010}, and it has been shown analytically only recently using CFT 
techniques~\cite{calabrese2012,calabrese2013a}. 
Furthermore, the negativity is attracting increasing attention in $D>1$ as an alternative 
indicator of topological order~\cite{lee2013,claudio2013}. 

\begin{figure}
\includegraphics[width=0.92\columnwidth]{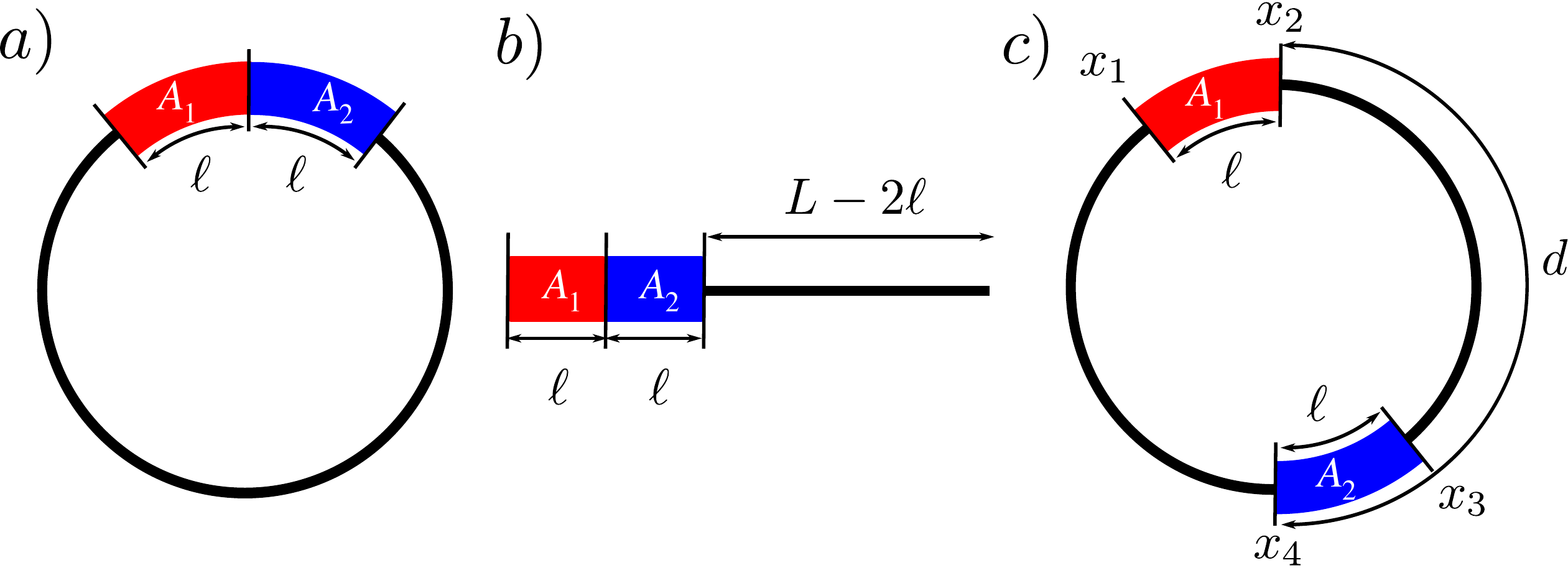}
\caption{Geometrical setup (chain partitions) 
 used in this work: two adjacent intervals ($A_1$, $A_2$)  of  
 equal length $\ell$ embedded in a chain (of length $L$) with 
 periodic (a) and open (b) boundary conditions. (c) Two 
 disjoint intervals.
}
\label{fig1:geometry}
\end{figure}

In this work we investigate the scaling behavior of $\tr(\rho_A^{T_2})^n$, i.e.
the $n-$th moment of $\rho_A^{T_2}$,
from which the negativity can in principle be obtained as the analytic continuation~\cite{calabrese2012,
calabrese2013a} ${\cal E}=\lim_{n\to 1}\tr(\rho_A^{T_2})^{n}$ ($n\in{\mathbb N}$ 
even).
Although not being proper entanglement 
measures, in 1D they provide universal information about critical systems. 
Specifically, for two adjacent intervals (cf. Fig.~\ref{fig1:geometry} (a)) 
their scaling behavior depends solely on the central charge, whereas for 
disjoint ones (Fig.~\ref{fig1:geometry} (c)) it can potentially reveal 
complete information about a CFT~\cite{calabrese2012,calabrese2013a}.

\paragraph*{Summary of results.---} We provide a novel Quantum Monte Carlo 
(QMC) scheme to calculate  the moments of the transposed density matrix, $\tr(\rho_A^{T_2})^n$, at  
finite temperature, using the replica trick (similarly to Monte Carlo 
approaches for the Renyi entropies~\cite{caraglio2008,alba2010,gliozzi2010,
alba2011,alba2013,hastings2010,melko2010,singh2011,isakov2011,kaul2012,
inglis2013,iaconis2013,humeniuk2012,chung13}). Our scheme generalizes that  proposed in Ref.~\onlinecite{alba2013} using  {\it classical} 
Monte Carlo. While only universal features can be accessed easily via 
classical simulations, in QMC both universal and non universal aspects can be accessed directly. For instance, temperature is a tunable parameter in QMC, whereas this 
is not possible, in an easy manner, within the classical Monte Carlo scheme. Thus, 
the QMC approach is ideal for benchmarking future finite temperature CFT results. 
Interestingly, it should also be possible  (in principle) to reconstruct the spectrum of 
$\rho_A^{T_2}$ (and hence the negativity), 
as we have demonstrated recently for the reduced density matrix in 
Ref.~\onlinecite{chung13} (see also Ref.~\onlinecite{assaad13}).

Instead of considering $\tr(\rho_A^{T_2})^n$, we introduce the combinations 
$r_n$ and $R_n$ (respectively for adjacent and disjoint intervals, see 
below for their definitions). These are scale invariant at a critical point and 
can be used as witnesses of exotic (topological) critical behaviors (similarly to 
Binder cumulants~\cite{binder1981} for standard criticality).
% as they do not rely on the concept of a local order parameter.  

To be specific, here we consider 1D hard-core bosons (at half and 
quarter filling) and the Bose-Hubbard chain in the superfluid phase. Both  
are special instances of the Luttinger liquid, which is a $c=1$ CFT. Surprisingly,   
for hard-core bosons, despite integrability, it is a formidable challenge 
to calculate analytically $\rho_A^{T_2}$ (in contrast with the case of free 
bosons~\cite{audenaert2002}).

At low enough temperature,  we find that $r_n$ is in excellent agreement with 
CFT~\cite{calabrese2012,calabrese2013a} for large enough chains, while at small 
sizes  {\it unusual} (in the sense of Ref.~\cite{cardy2010b}) corrections are 
present. These arise from the {\it local} breaking of conformal invariance near 
the endpoints of the intervals, and are generic for entanglement-related 
quantities and show parity oscillations alike standard Renyi entropies~\cite{nienhuis2009,calabrese2010a,calabrese2010b}.
Detailed knowledge of these corrections is imperative with respect to the application of entanglement 
related tools as indicators of critical behavior.
% Their  amplitude oscillates as a function of the interval size 
% (parity effects) with a filling-dependent period, as it has been found for 
% standard entropies~\cite{nienhuis2009,calabrese2010a,calabrese2010b}.
% Besides being interesting {\it per se} due to their intriguing origin, 
% their knowledge is necessary in view of the applications of entanglement 
% related tools as indicators of critical behavior. 

% For disjoint intervals, unusual corrections are in general stronger. For 
% hard-care bosons we extract both their exponent $\omega'_n$ and amplitude. In 
% particular, we provide convincing evidence that $\omega'_n=2/n$, which suggests 
% that $\omega'_n=2K_L/n$ in a generic Luttinger liquid (here $K_L$ is the Luttinger 
% parameter). This is  the same exponent governing the corrections of the mutual 
% information between two intervals~\cite{alba2010,alba2011}. 
% Finally, in the Bose-Hubbard chain scaling corrections are smaller and we find perfect agreement 
% with CFT predictions already at finite but large enough lattices. 

For hard-care bosons we provide convincing evidence that the leading exponent of these corrections in the two-interval case -- where unusual contributions are generally stronger -- is $\omega'_n=2/n$.
This suggests that $\omega'_n=2K_L/n$ in a generic Luttinger liquid (here $K_L$ is the Luttinger 
parameter) wich is the same as for the mutual information~\cite{alba2010,alba2011}.
Finally, in the Bose-Hubbard chain scaling corrections are smaller and we find perfect agreement 
with CFT predictions already at finite but large enough lattices. 

%%%%%%%%%%%%%%%%%%%%%%%%%%%%%%%%%%%%%%%%%%%%%%%%%%%%%%
%%%%%%%%%%%%%%%%%%%%%%%%%%%%%%%%%%%%%%%%%%%%%%%%%%%%%%

\paragraph*{Models \& observables.---} We mainly consider the (integrable) 1D 
hard-core boson (at most one particle per lattice site) model with $L$  sites, which is defined by the Hamiltonian 
${\mathcal H} = -t\sum_{i} (b_i^\dagger b_{i+1} +h.c.)$. 
Here $b_i$ are bosonic annihilation operators and $t=1$ is the hopping amplitude. 
% The periodic chain is obtained connecting sites $1$ and $L$. 
% The on-site particle number satisfies the constraint $n_i\equiv b^{\dagger}_ib_i=0,1\, \forall i$.
% In particular we work at non-integer (half and quarter) fillings.
In particular we work at half and quarter fillings.
We also consider the 1D Bose-Hubbard model given by the Hamiltonian ${\mathcal H} = -t\sum_{i=1}
(b_i^\dagger b_{i+1} +h.c.) + U/2 \sum_{i} n_i ( n_i-1)$, where $U$ is the interaction strength. 
Specifically we restrict ourselves to the superfluid phase at unit filling 
and fix $U=2$ (the Mott-superfuid transition being at $U\approx 3.38$~\cite{kuhner1998}). 
% In both cases the spectrum of the model is gapless, and its low-energy properties are  described by the Luttinger liquid theory. 
The low energy properties of both models are captured by a gapless Luttinger liquid.
For hard-core boson model the Luttinger parameters $K_L=1$, while for Bose-Hubbard model 
$K_L\approx 3.125$ at $U=2$~\cite{rachel2012,lauchli2013}.
For two adjacent intervals [cf. Fig.~\ref{fig1:geometry}(a),(b)] we define 
(following~\cite{calabrese2012}) the ratios $r_n$ ($n\in{\mathbb N}$)
\begin{equation}
\label{rn}
r_n(z)\equiv\log\left[\frac{\tr(\rho^{T_2=\ell}_{A_1\cup A_2})^n}
{\tr(\rho^{T_2=L/4}_{A_1\cup A_2})^n}\right],
\end{equation}
with $z\equiv\ell/L$~\footnote{Choosing $\ell=L/4$ in the denominator of the definition of $r_n(z)$ ensures that non-universal contributions cancel in the thermodynamic limit~\cite{calabrese2012}. This covention leads, for instance, to a crossing of $r_3$ at $z=1/4$ (see Fig.~\ref{fig3:r3}) that must not be taken as an indicator of universality at this specific value of $z$.}. 
Here the notation $\rho^{T_2=\ell}$ means that the 
partial transposition is done with respect to the degrees of freedom of 
subsystem $A_2$ of length $\ell$ (see Fig.~\ref{fig1:geometry} 
(a)(b)). For two disjoint intervals it is convenient to define $R_n$ as  
\begin{equation}
\label{Rn}
R_n(y)\equiv \frac{\tr(\rho^{T_2}_{A_1\cup A_2})^n}
{\tr\rho_{A_1\cup A_2}^n},
\end{equation}
where $y$ is the four point ratio $y\equiv|(x_2-x_1)(x_4-x_3)|/
|(x_3-x_1)(x_4-x_2)|$ [see Fig.~\ref{fig1:geometry}(c)],  and one has 
$|x_i-x_j|\to L/\pi\sin(\pi|x_i-x_j|/L)$ (chord length) for finite chains. 
By construction, all the length scales and non universal factors cancel in 
Eqs.~\eqref{rn} and \eqref{Rn}. As a consequence 
$r_n(z)$ and $R_n(y)$ are scale invariant quantities (for any $n,z,y$) 
at criticality apart from scaling corrections at 
finite $L,\ell$. 
Moreover, while $r_n(z)$ depends only on the central charge, 
much more universal information is contained in $R_n(y)$.

%%%%%%%%%%%%%%%%%%%%%%%%%%%%%%%%%
\begin{figure}
\includegraphics[width=0.95\columnwidth]{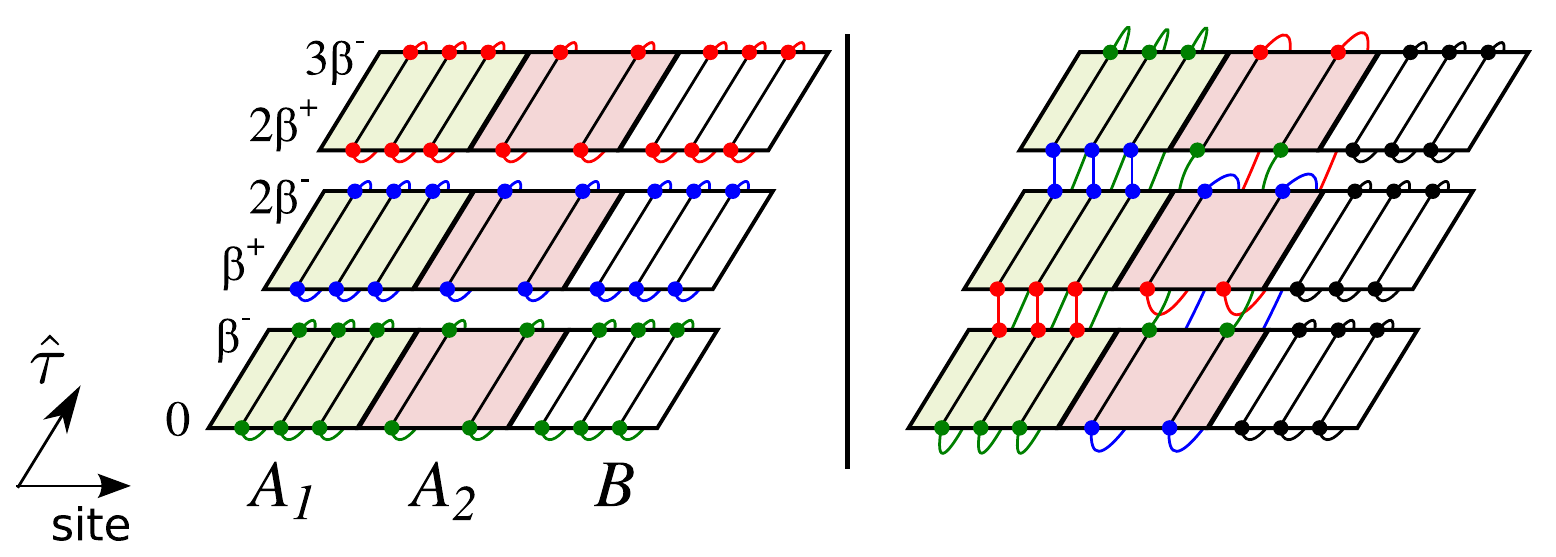}
\caption{Replica trick scheme for calculating 
 the $n$th moment (here $n=3$) of $\rho_A^{T_2}$ in Quantum Monte 
 Carlo (QMC) simulations. Left: three disconnected replicas (of area $L\times 
 \beta$). On each replica periodic boundary conditions are used along the 
 imaginary time direction $\hat\tau$. Right: Topology of  ${\mathcal K}_3$ 
 for two adjacent intervals. Colored links now connect points on 
 different replicas. Regions corresponding to different intervals 
 $A_1,A_2$ (see Fig.~\ref{fig1:geometry}) are shaded with different 
 colors.
}
\label{fig2:wld}
\end{figure}
%%%%%%%%%%%%%%%%%%%%%%%%%%%%%%%%%

%#####
\paragraph*{The moments of $\rho_A^{T_2}$: QMC algorithm.---}
The moments of the partially transposed reduced density matrix 
$\tr(\rho_A^{T_2})^n$ can be measured in Quantum Monte Carlo 
(QMC) simulations by exploiting a suitable replica representation.  
Given a generic lattice model, one has~\cite{calabrese2012}
\begin{equation}
\label{replica_rep}
\tr(\rho_{A_1\cup A_2}^{T_2})^n =\frac{Z_n^{T_2}(A_1\cup A_2)}{Z^n}, 
\end{equation}
where $Z=\tr\exp(-\beta H)$ is the partition function at temperature 
$T=1/\beta$, while $Z_n^{T_2}$ is defined  over an {\it ad hoc} 
surface ${\mathcal K}_n$, obtained by ``gluing'' $n$ disconnected replicas. 
For $n=3$ and two adjacent intervals ${\mathcal K}_n$ is illustrated  
in Fig.~\ref{fig2:wld} (right), and is formally obtained by introducing 
an equal-time branch cut (lying along subsystem $A$) at $\tau=k\beta, 
k=1,2,\dots,n$ on each replica. Links crossing the branch cuts (colored 
links in Fig.~\ref{fig2:wld} (right)) connect sites on different replicas. 
The ``gluing'' scheme is different for the two intervals $A_1,A_2$, reflecting 
the partial transposition on $A_2$. 

The ratio in Eq.~\eqref{replica_rep} can be sampled using a world-line 
based QMC. Here we use a continuous time world algorithm~\cite{prokofev98a,
prokofev98b,Pollet2007} (extensions to other QMC schemes  are straightforward),  
supplementing the standard world line update  with a  non-local move.   
Given that the system is on $\mathcal{K}_n$ [cf. Fig.~\ref{fig2:wld} (right)], 
the move tries to cut all the world lines at  $k\beta^{+}$ and $k\beta^{-}$, 
creating new ones connecting sites at $k\beta^{+}$ and $(k+1)\beta^{-}$, 
as in Fig.~\ref{fig2:wld} (left) (note the periodicity in imaginary time).
If the move is possible, the global topology is changed from 
$\mathcal{K}_n$ to $n$ disconnected sheets.
The inverse move from 
$n$ disconnected sheets to $\mathcal{K}_n$ is performed in a similar fashion.
Finally, one measures $\tr(\rho_A^{T_2})^n=\langle N^{c}_A/N^{dis}_{A}
\rangle$, where $N^{c}_{A}$ and $N^{dis}_A$  are the total number of 
QMC steps happening on the connected replicas ${\mathcal K}_n$
and disconnected sheets respectively, and $\langle\cdot\rangle$ is the Monte Carlo average.

Upon increasing the length of $A$ as well as the replica index $n$ the transition 
probabilities in the global update become small, severely limiting the 
performance of the algorithm. 
To circumvent these issues we use the so called \textit{increment tricks}~\cite{hastings2010,chung13}.
\paragraph*{Two adjacent intervals.---}

%%%%%%%%%%%%%%%%%%%%%%%%%%%%%%%%%
\begin{figure}
\includegraphics[width=0.93\columnwidth]{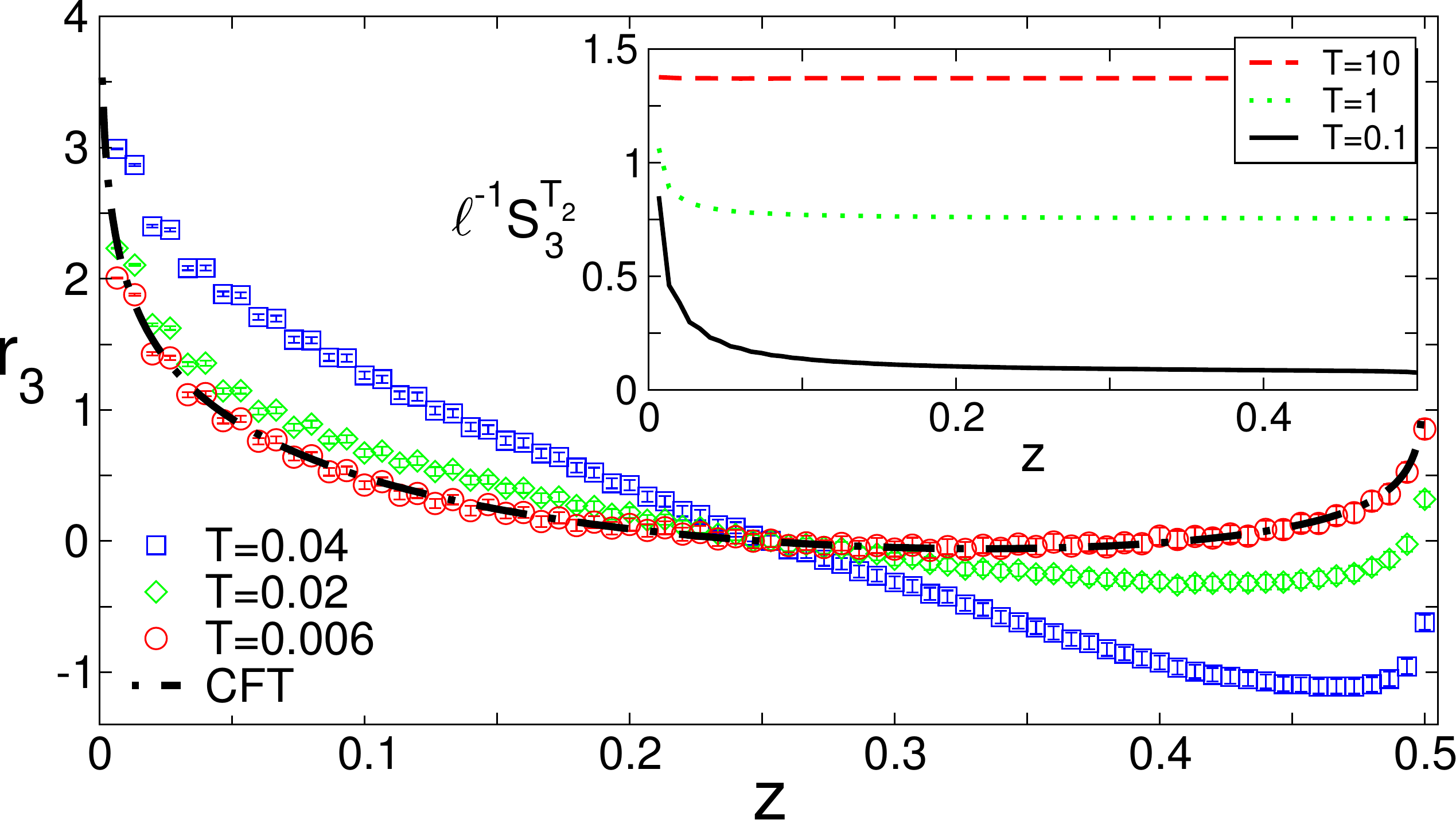}
\caption{
 %Two adjacent intervals: scale invariant ratio $r_3$ for  
 %a periodic chain (with $L=150$ sites) of hard-core bosons at 
 %half filling. QMC data for $r_3$ versus   $z\equiv\ell/L$, 
 %with $\ell$ the being length of the interval and for several 
 %temperatures $T$ is shown. 
 %The dashed-dotted line  is the zero temperature 
 %CFT result. Note that the crossing at $z=1/4$ is merely due to the 
 %definition of $r_n$. Inset: $\ell^{-1}S_3^{T_2}$ ($S_3^{T_2}
 %\equiv-\log\tr(\rho_A^{T_2})^3$) versus $z$ (QMC data). 
 %Note the volume law already at $T\sim 1$. 
Two adjacent intervals: We show QMC data for $r_3$ versus   $z\equiv\ell/L$ for a periodic chain of length $L=150$ of hard-core bosons at half filling for different temperatures.
 The dashed-dotted line  is the zero temperature CFT result.
 Note that the crossing at $z=1/4$ is merely due to the definition of $r_n$.
 Inset: $\ell^{-1}S_3^{T_2}$ versus $z$ (QMC data) showing that the transposed Renyi entropy exhibits a volume law already at $T \sim 1$.
}
\label{fig3:r3}
\end{figure}
%%%%%%%%%%%%%%%%%%%%%%%%%%%%%%%%%

As a benchmark of the algorithm we first focus on two adjacent (equal-size) 
intervals [cf. Fig.~\ref{fig1:geometry} (a,b)], discussing the scaling 
invariant ratios $r_n$ ($n=3,4$). Fig.~\ref{fig3:r3} 
plots $r_3$ as function of $z\equiv \ell/L$ (data for a periodic  
hard-core boson chain of length $L=150$ and several temperatures).  
At $T=0$, $r_n(z)$ (for any $n$) can be obtained analytically using Eq.~\eqref{rn} 
and in any CFT one has~\cite{calabrese2012,calabrese2013a,calabrese2013b}
\begin{align}
& \nonumber\tr(\rho_A^{T_2})^{n_e} \propto (\ell_1\ell_2)^{-\frac{c}{6}
(\frac{n_e}{2}-\frac{2}{n_e})}(\ell_1+\ell_2)^{-\frac{c}{6}(\frac{n_e}{2}+
\frac{1}{n_e})}\\
& \tr(\rho_A^{T_2})^{n_o} \propto (\ell_1\ell_2(\ell_1+\ell_2))^{-
\frac{c}{12}(n_o-\frac{1}{n_o})},
\label{adjacent_scal}
\end{align}
with $\ell_i$ the two intervals lengths, $n_e$($n_o$) an even(odd) integer, 
and $c$ the central charge. The resulting theoretical curve (after replacing 
$\ell_i\to L/\pi\sin(\pi\ell_i/L)$ in Eq.~\eqref{adjacent_scal}) is 
plotted in Fig.~\ref{fig3:r3} as a dashed-dotted line. At $T=0.006$, QMC data 
perfectly agree with CFT (i.e.~scaling corrections are small). Interestingly, 
$r_3$ provides an effective way of extracting $c$. Indeed, fitting QMC data 
to Eq.~\eqref{adjacent_scal}, one obtains $c=0.98(5)$, fully compatible 
with $c=1$. On the other hand, finite temperature effects are already visible at $T=0.02$.

%%%%%%%%%%%%%%%%%%%%%%%%%%%%%%%%%
\begin{figure}
\includegraphics[width=0.99\columnwidth]{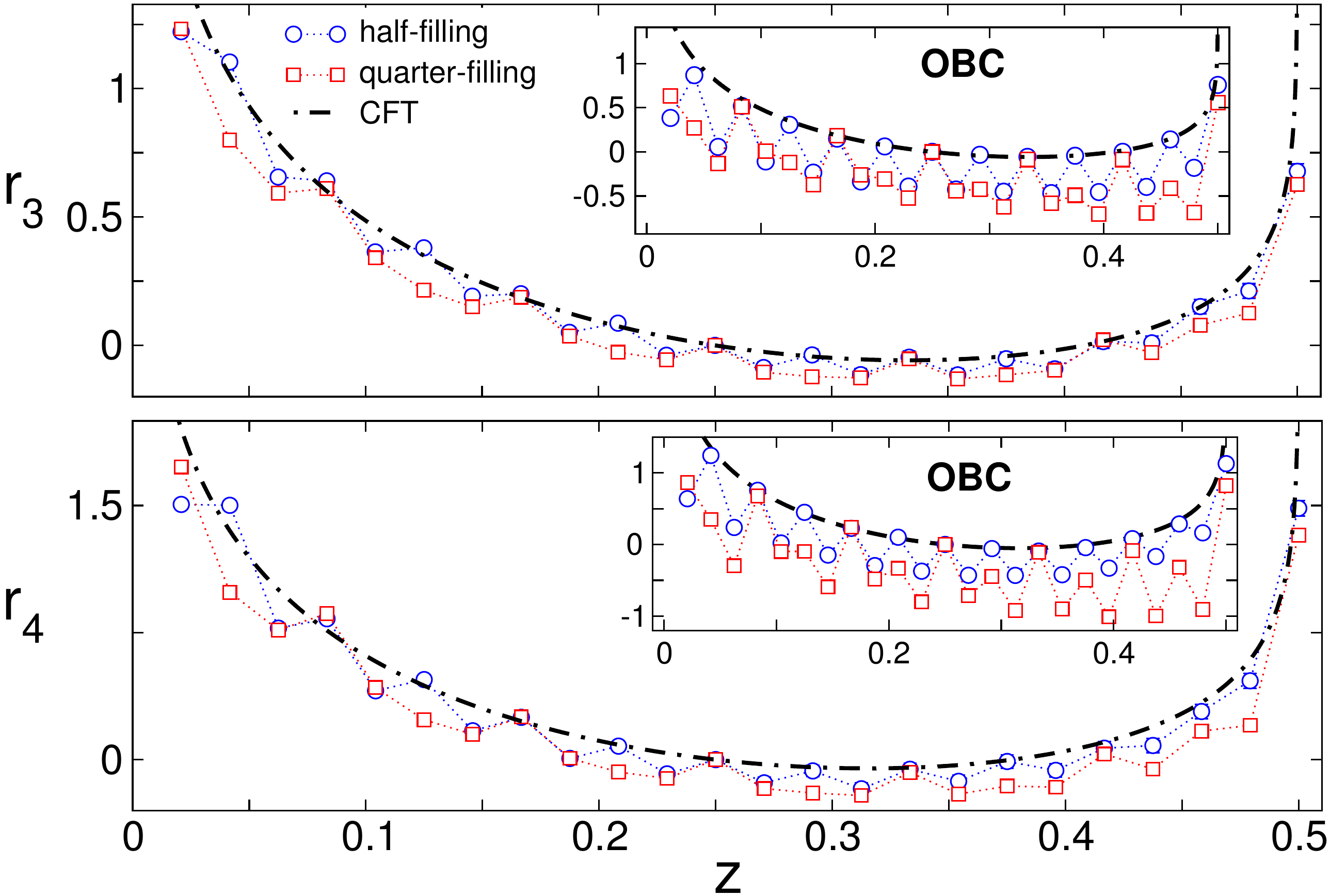}
\caption{Scaling corrections of $r_3$ and $r_4$: parity and boundary 
 effects. $r_3$ (top) and $r_4$ (bottom) from QMC as 
 function of $z\equiv\ell/L$ at fixed  $L=48$ and $T=0.01$ for a periodic chain of hard-core bosons at half (circles) 
 and quarter (squares) filling. 
 %QMC data for a periodic chain of hard-core bosons at half (circles) 
 %and quarter (squares) filling. 
 The dashed-dotted line is the zero 
 temperature CFT result (same as in Fig.~\ref{fig3:r3}). Inset: 
 same as in the main figure but for open boundary conditions. The 
 amplitude of the corrections is enhanced. 
}
\label{fig4:r3open}
\end{figure}
%%%%%%%%%%%%%%%%%%%%%%%%%%%%%%%%%

It is also instructive to consider the ``transposed entropy'' $S_n^{T_2}
\equiv-\log\tr(\rho_A^{T_2})^n$ (see the inset in Fig.~\ref{fig3:r3}). 
At $T\to\infty$ one expects (for the infinite chain) a thermodynamic volume law 
$S_n=S_n^{T_2}=2\ell\log 2,\forall n$. This is already visible at $T\sim 10$
reflecting that $\rho_A$ is almost diagonal. Since at high temperature 
only classical correlations survive, this implies that $S_n^{T_2}$ is not a good 
entanglement measure.

%%%%%%%%%%%%%%%%%%%%%
\paragraph*{Unusual scaling corrections.---}

One intriguing feature of entanglement related quantities is that they 
exhibit unusual finite size scaling corrections~\cite{cardy2010b}. These 
arise from conical singularities near the endpoints of the subsystems, 
and can depend on both irrelevant and {\it relevant} operators (in the 
renormalization group sense) of the theory, whereas usual corrections 
are due only to irrelevant ones. 

% For the Renyi entropies $S_n$, in a system described by a Luttinger liquid, 
% unusual corrections oscillate with the parity of $\ell$, and can be given 
% as~\cite{calabrese2010a,calabrese2010b}
% %
% \begin{equation}
% \label{scal_ansatz}
% S_n(\ell)-S_n^{CFT}(\ell)= f_n\cos(2k_F\ell)\ell^{-\omega_n}, 
% \end{equation}
% %
% with $f_n$ a nonuniversal amplitude and $k_F$ the Fermi momentum. Interestingly, 
% $\omega_n$ depends on the Luttinger parameter $K_L$, as $\omega_n=2K_L/n$, 
% and on global geometric properties  (cf. Fig.~\ref{fig2:wld}), through 
% the Renyi index $n$. In presence of boundaries (open boundary conditions) 
% Eq.~\eqref{scal_ansatz} holds true replacing $\omega_n\to K_L/n$.

For Luttinger liquid unusual corrections lead to parity oscillations of the Renyi entropies that can be given 
as~\cite{calabrese2010a,calabrese2010b}
\begin{equation}
\label{scal_ansatz}
S_n(\ell)-S_n^{CFT}(\ell)= f_n\cos(2k_F\ell)\ell^{-\omega_n}, 
\end{equation}
with $f_n$ a nonuniversal amplitude and $k_F$ the Fermi momentum. 
Notably, $\omega_n$ depends on the Luttinger parameter $K_L$ as well as on the Renyi index (and thus on the global geometry) as $\omega_n=(2)K_L/n$ for the open (periodic) case.
% Interestingly, $\omega_n$ depends on the Luttinger parameter $K_L$, as $\omega_n=2K_L/n$, 
% and on global geometric properties  (cf. Fig.~\ref{fig2:wld}), through 
% the Renyi index $n$. In presence of boundaries (open boundary conditions) 
% Eq.~\eqref{scal_ansatz} holds true replacing $\omega_n\to K_L/n$.

It is natural to expect similar corrections for $S_n^{T_2}$ (and for $r_n(z)$ 
thereof).
% , implying that Eq.~\eqref{scal_ansatz}  holds (with different $f_n$). 
This is supported in Fig.~\ref{fig4:r3open} plotting QMC data for 
$r_3$ and $r_4$ as function of $z$ showing data for both periodic and open boundary 
conditions at fixed $L=48$,
and half and quarter filling ($k_F=\pi/2$ and $k_F=\pi/4$). Clearly, scaling corrections oscillate consistently with $\sim\cos 2k_F\ell$ 
(for both open and periodic boundary conditions), in agreement with a generalization of
Eq.~\eqref{scal_ansatz} to the case of two intervals. Interestingly, as for the standard   
entropies~\cite{laflorencie2006,affleck2009}, the corrections amplitude   
is enhanced (for $r_3$ by a factor $\sim 10$) with open boundary 
conditions (cf. insets in Fig.~\ref{fig4:r3open}).

%%%%%%%%%%%%%%%%%%%%%%%%%%%%%%%%%%%%%%%%%%%%%%%
\begin{figure}[t]
\includegraphics[width=0.92\columnwidth]{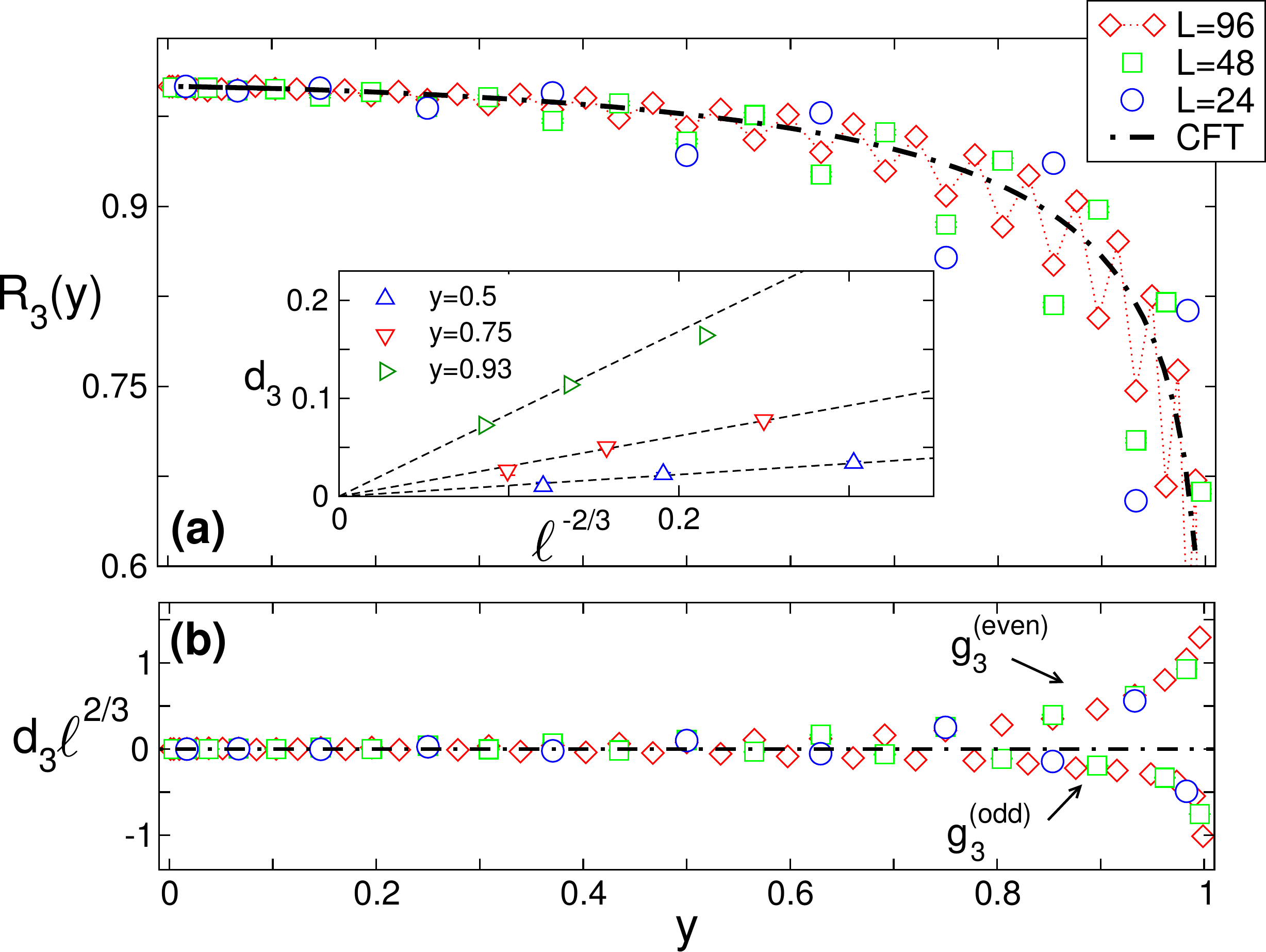}
\caption{1D hard-core bosons: the ratio $R_3(y)$. 
 QMC data at half-filling, chain lengths $L=24,48,96$ (periodic 
 boundary conditions), and temperatures $T=0.24/L$. (a) $R_3(y)$ 
 vs cross ratio $y$. The dotted line highlights  
 the oscillating behavior. The dashed-dotted line is the 
 zero temperature CFT result. Inset: $d_3\equiv 
 R^{CFT}_3-R_3$  (at fixed $y$) versus $\ell^{-2/3}$, 
 $\ell$ being the intervals size. Dashed lines are one parameter 
 fits to $\sim\ell^{-2/3}$. (b) Amplitude  
 $d_3 \ell^{2/3}$ of the corrections plotted versus $y$. 
 Note the data collapse with the functions $g_3^{(q)}(y)$, 
 $q$ being the parity of $\ell$. 
}
\label{fig5:R3}
\end{figure}
%%%%%%%%%%%%%%%%%%%%%%%%%%%%%%%%%%%%%%%%%%%%%%%

%%%%%%%%%%%%%%%%%%%%%%
\paragraph*{Two disjoint intervals.---}
We now turn to the more complicated situation  of two disjoint intervals in 
a periodic chain (see Fig.~\ref{fig1:geometry} (c)), focusing on the ratio 
$R_3(y)$ (see Eq.~\eqref{Rn}). In the asymptotic limit (after sending all 
the length scales to infinity), for any model described by a CFT it is given as~\cite{calabrese2012,calabrese2013a} 
\begin{equation}
\label{Rn_CFT}
R^{CFT}_n(y) =(1-y)^{\frac{c}{3}(n-\frac{1}{n})}\frac{
\mathcal{F}_n(y/(y-1))}{\mathcal{F}_n(y)}
\end{equation}
with $y$ the four point ratio, $c$ the central charge, and ${\mathcal F}_n(x)$ 
a {\it universal} scaling function, which depends on the full operator content of 
the underlying CFT. The analytical form of ${\mathcal F}_n(x)$ is known  
only for the Luttinger liquid and the 1D Ising universality class (see Refs.~\citep{
alba2010,calabrese2009a,calabrese2011} for their precise expression). 

$R_3(y)$ versus $y$ for hard-core bosons at quarter and half filling is shown in Fig.~\ref{fig5:R3} (a) 
% (QMC data for  hard-core 
% bosons at half-filling). 
Different values of $y$ on $x$-axis are obtained by varying the 
length of the two intervals at fixed $d=L/2$ (cf. Fig.~\ref{fig1:geometry} 
(c)). The dashed-dotted line is the asymptotic CFT result from 
Eq.~\eqref{Rn_CFT}.  In the limit $y\to 0$, i.e. two intervals far apart 
($d-\ell\to\infty$ in Fig.~\ref{fig1:geometry}), one has $\rho_{A_1\cup A_2}
\approx\rho_{A_1}\otimes\rho_{A_2}$, implying $R_n\to 1$. 
Oppositely, at $y\to 1$ the case of two adjacent intervals is recovered, and 
from Eq.~\eqref{adjacent_scal}  one has $R_n\to 0$. 

For finite chains we find oscillating corrections that are similar to those of the mutual 
information between two disjoint intervals~\cite{alba2010,fagotti2010,alba2011,
maurizio2011,fagotti2011,fagotti2012}). Under general assumptions, for any $n$ 
their behavior can be given as 
\begin{equation}
\label{corr}
R_n(y)=R_n^{CFT}(y)+ g^{(q)}_n(y)\ell^{-\omega'_n}+\dots,
\end{equation}
with $\omega'_n$ being the corrections exponent, and $g^{(q)}_n(y)$ their    
amplitude, which depends on both $y$ and the parity $q$ of the interval 
length (the dots in Eq.~\eqref{corr} denote more irrelevant terms).  
%
%%%%%%%%%%%%%%%%%%%%%%%%%%%%%%%%%%%%%%%%%%%%%%%
\begin{figure}[t]
\includegraphics[width=0.95\columnwidth]{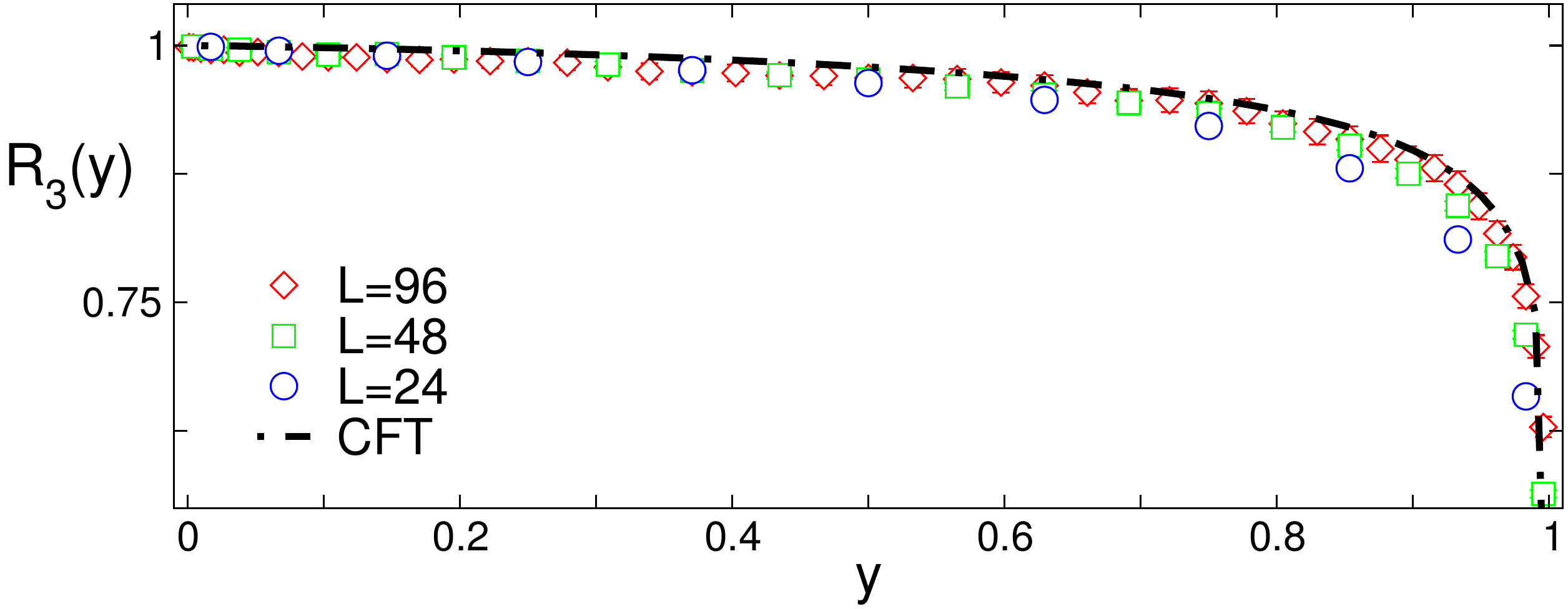}
\caption{Bose-Hubbard chain in the superfluid phase: scale invariant ratio 
 $R_3(y)$ versus the cross ratio $y$. QMC data are for a periodic chain at 
 $U=2$ (corresponding to Luttinger parameter 
 $K_L\approx 3.125$), and temperatures $T=0.96/L$. The dashed-dotted 
 line is the asymptotic zero temperature CFT result.
}
\label{fig6:R3BH}
\end{figure}
%%%%%%%%%%%%%%%%%%%%%%%%%%%%%%%%%%%%%%%%%%%%%%%
%
A standard finite size scaling analysis, fitting QMC data at fixed $y=1/2$ 
to $\sim1/\ell^{\omega'_3}$, gives $\omega'_3=0.6(1)$, which is consistent 
with $\omega'_3=\omega_3=2K_L/3=2/3$ (see Eq.~\eqref{scal_ansatz}).   
This is further supported in Fig.~\ref{fig5:R3} (inset) plotting 
$d_3\equiv R_3^{CFT}-R_3$ at fixed $y$ versus $\ell^{-2/3}$. Dashed 
lines are one parameter fits to $\sim\ell^{-2/3}$.  
%Our results suggest that $\omega'_n=\omega_n=2K_L/n$ holds true also away from the hard-core limit. 
The amplitudes $g^{(q)}_3(y)$ (extracted as $g_3^{(q)}\equiv 
d_3\ell^{2/3}$) are shown in Fig.~\ref{fig5:R3} (b), and are rapidly vanishing 
at $y\to 0$. Also, data for different sizes collapse on the two curves, 
confirming {\it a posteriori} the correctness of our analysis.

% Finally, to demonstrate the versatility of our QMC method, in  Fig.~\ref{fig5:R3BH}  we discuss $R_3(y)$ for the Bose-Hubbard (periodic) chain in the superfluid phase. 
Finally, to demonstrate the versatility of our QMC approach, we discuss $R_3(y)$ in the superfluid Bose-Hubbard model on a periodic chain.
Already at $L=96$ QMC data are in 
impressive agreement (at any $0\le y\le 1$) with the asymptotic CFT result 
(dashed-dotted line in the Fig.~\ref{fig6:R3BH}). This  confirms that scaling corrections 
to $R_n(y)$ become smaller upon increasing the Luttinger parameter, (similarly 
to what has been observed in Ref.~\onlinecite{alba2013}).

%%%%%%%%%%%%%%%%%%%%
\paragraph*{Summary \& discussion.---}

We presented a Quantum Monte Carlo scheme for calculating the moments 
of the partially transposed reduced density matrix, both at zero and finite 
temperature. These are the main ingredients in CFT to calculate the 
logarithmic negativity. We considered several combinations ($r_n,R_n$) of 
the moments that are scale invariant at a 1D quantum critical point, and, not 
relying on any local order parameter, could be useful to detect exotic (topological) 
critical behavior. After taking into account unusual (oscillating) scaling 
corrections, their behavior is in full agreement with recent CFT results,  
for both 1D hard-core bosons and the 1D Bose-Hubbard model. 

Our results pave the way to many possible research directions. First, 
it would be interesting to apply the method to higher dimensions, especially  
to investigate the behavior of $r_n,R_n$ in topologically ordered phases of 
matter. 

Also, it should be possible to obtain high-temperature series expansions 
(in any dimension) for the moments of $\rho_A^{T_2}$ and the negativity itself. 
Our QMC scheme could then be used as a useful benchmark method.

\paragraph*{Acknowledgements.---} We acknowledge support by the Austrian 
Science Fund (FWF) through the SFB FoQuS (FWF Project No.~F4018-N23). 
We acknowledge financial support and allocation of CPU time from NSC and 
NCTS Taiwan. This work was supported by the Austrian Ministry of Science 
BMWF as part of the UniInfrastrukturprogramm of the Forschungsplattform 
Scientific Computing at LFU Innsbruck.

%\bibliographystyle{apsrev4-1.bst}
%\bibliography{negativity}
%
\end{document}